\documentclass[preprint,aps]{revtex4}
\usepackage{graphicx}

\newcommand{\pr}{^{\prime}}

\newcommand{\be}{\begin{eqnarray}}
\newcommand{\ee}{\end{eqnarray}}
\newcommand{\la}{\langle}
\newcommand{\ra}{\rangle}

\begin{document}
\title{Zemach and magnetic radius of the proton from \\
the hyperfine splitting in hydrogen}
\author{A. V. Volotka,$^{1,2}$ V. M. Shabaev,$^{1,2}$ G. Plunien,$^2$
and G. Soff$^2$}

\affiliation{
$^1$Department of Physics, St.Petersburg State University,
Oulianovskaya 1, Petrodvorets, St.Petersburg 198504, Russia\\
$^2$Institut f\"ur Theoretische Physik, TU Dresden,
Mommsenstra{\ss}e 13, D-01062 Dresden, Germany}

\begin{abstract}
The current status of the determination of corrections to the hyperfine
splitting of the ground state in hydrogen is considered. Improved calculations
are provided taking into account the most recent value for
the proton charge radius. Comparing experimental data with predictions for
the hyperfine splitting, the Zemach radius of the proton is deduced to be
$1.045(16)$ fm.
Employing exponential parametrizations for the electromagnetic form
factors we determine the magnetic radius of the proton to be
$0.778(29)$ fm.
Both values are compared with the corresponding ones derived from the data
obtained in electron-proton scattering experiments and the data extracted
from a rescaled difference between the hyperfine splittings in hydrogen and
muonium.
\end{abstract}
\maketitle

\section{Introduction}

High-precision measurements and calculations of energy spectra of hydrogen-like
atoms provide tests of quantum electrodynamics (QED) with very high precision  (see
\cite{udem1997,schwob1999,beauvoir2000,pachucki2003} and references therein). In
some cases, the current
accuracy of QED calculations exceeds those of the known values of fundamental
physical constants. For instance, recent measurements and calculations of the $g$ factor
of hydrogen-like carbon and oxygen have provided the basis for a new determination of
the electron mass (see \cite{verdu2004} and references therein). Measurements of the Lamb
shift in hydrogen, combined with corresponding calculations, have facilitated to
determine the Rydberg constant and to deduce an improved value for the proton charge
radius \cite{mohr2004,melnikov2000,pachucki2001,karshenboim2003}.

The relative experimental accuracy of the ground-state hyperfine splitting in hydrogen
is better than $10^{-12}$ \cite{hellwig1970}. The error associated with the QED corrections
to the hyperfine splitting is estimated to contribute on the level $10^{-9}$. The
major theoretical uncertainty arises from nuclear structure-dependent contributions.
The most important structure-dependent term is the proton-size correction, which is
determined exclusively by the spatial distributions of the charge and the magnetic
moment of the proton. It contributes on the relative level $10^{-5}$. Assuming that
all other theoretical corrections are accurately known, one can determine the
proton-size contribution by comparing theoretical and experimental values for the
hyperfine splitting in hydrogen. The major goal of the present paper is to determine
the Zemach and the magnetic radius of the proton by such a comparison.

In Sec. II, we consider various theoretical contributions to the hyperfine splitting
and derive the  proton-size correction comparing theory and experiment. In Sec. III,
we refine the value of the  proton-size correction by recalculating some of these
contributions and determine the magnetic radius of the proton, employing an exponential
parametrization for the electric and magnetic form factors. The recalculation of the
recoil correction has improved the value of the Zemach radius compared to the
previous result obtained in \cite{dupays2003}. In Sec. IV, the results obtained are
compared with corresponding data derived from elastic electron-proton scattering
experiments \cite{sick2003,friar2004,hammer2004} and the data extracted from a rescaled
difference between the hyperfine splittings in hydrogen and muonium \cite{brodsky2004}.

The relativistic units $\hbar=c=1$ are used throughout the paper.

\section{Hyperfine splitting in hydrogen}

The hyperfine splitting of the ground state in hydrogen can be written in the form
\be\label{hydrogen}
\Delta E_{{\rm theor}}=E_{{\rm F}}
(1+\delta^{{\rm Dirac}}+\delta^{{\rm QED}}+\delta^{{\rm structure}})\,,
\ee
where $E_{{\rm F}}$ is the Fermi energy \cite{fermi1930}
\be
E_{{\rm F}}=\frac83\alpha(\alpha Z)^3
\frac{m_{\rm e}^2 m_{\rm p}^2}{(m_{\rm e}+m_{\rm p})^3}\
\frac{\mu_{\rm p}}{\mu_{\rm N}}\,,
\ee
$\mu_{\rm p}$ is the magnetic dipole moment of the proton,
$\mu_{\rm N}$ is the nuclear magneton,
$m_{\rm e}$ and $m_{\rm p}$ are the electron and proton mass, respectively.
The relativistic correction $\delta^{{\rm Dirac}}$ can easily be
obtained from the Dirac equation \cite{breit1930}:
\be
\delta^{{\rm Dirac}}=\frac32(\alpha Z)^2+\frac{17}{8}(\alpha Z)^4+...\,.
\ee
Here and in what follows we keep
the nuclear charge number $Z$ to separate the
relativistic  and radiative corrections.
For recent achievements in calculations of
the radiative correction
$\delta^{{\rm QED}}$ we refer to Refs.
\cite{pachucki1996,kinoshita1996,kinoshita1998,karshenboim2000,yerokhin2001,karshenboim2002}.
The uncertainty of
$\delta^{{\rm QED}}$ is mainly determined by  uncalculated
terms of order $\alpha^3(\alpha Z)$ and by uncertainties associated with some
of the calculated terms. The structure-dependent correction
$\delta^{{\rm structure}}$ is usually expressed as the sum
\be
\delta^{{\rm structure}}=\delta^{{\rm pol}}
+\delta^{\mu{\rm vp}}+\delta^{{\rm hvp}}+\delta^{{\rm weak}}+\delta^{{\rm rigid}}\,.
\ee
The part associated with intrinsic proton dynamics (polarizability)
$\delta^{{\rm pol}}$ has been recently evaluated in
\cite{faustov2002} employing experimental and theoretical results
for the structure functions of polarized protons.
The correction due to muonic vacuum-polarization $\delta^{\mu{\rm vp}}$
has been obtained in \cite{karshenboim1997}, while the hadronic
vacuum-polarization contribution $\delta^{{\rm hvp}}$ was
evaluated in \cite{faustov1999,martynenko2004}. For the weak interaction term
$\delta^{{\rm weak}}$ we refer to Refs. \cite{beg1974,eides1996}.
Values  for all these corrections together with corresponding
uncertainties are presented in Table I.
The Fermi energy $E_{{\rm F}}$ is evaluated  employing  the values of
the fundamental constants tabulated in \cite{mohr2004}.
The leading chiral logarithms contributions to the structure-dependent
correction have been also investigated within an effective field theory
\cite{pineda2003}.

Now let us turn to the term $\delta^{{\rm rigid}}$, which is determined
by electric and magnetic form factors of the proton. This quantity
can be decomposed into two parts:
$\delta^{{\rm rigid}}=\delta^{{\rm ps}}+\delta^{{\rm recoil}}$,
where $\delta^{{\rm ps}}$ represents the  proton-size correction and
$\delta^{{\rm recoil}}$ is
associated with recoil effects.
The recoil part contains both terms arised from a pointlike Dirac
proton and additional recoil correction due to the internal proton
structure. Following Ref. \cite{bodwin1988} we do not separate them.
Calculations of the
dominant contribution (relative order $(\alpha Z)m_{{\rm e}}/m_{{\rm p}}$)
to the recoil correction have a long history (see \cite{bodwin1988} and
references therein). The contribution of the order
$(\alpha Z)^2 m_{{\rm e}}/m_{{\rm p}}$ has been
first derived in \cite{bodwin1988}, while the radiative-recoil correction
of the order $\alpha(\alpha Z)m_{{\rm e}}/m_{{\rm p}}$ has been
obtained in \cite{karshenboim1997}.
To determine the magnetic radius of the proton from the hydrogen
hyperfine splitting we propose the following. At first we
calculate the structure-dependent part of the recoil correction
in a rough approach, taking the proton magnetic radius to be the same as
the charge one. Then we find the proton-size correction from a comparison
of the experimental and theoretical values of the hyperfine splitting.
Using the dipole parametrizations of the form factors we extract
a preliminary value for the proton magnetic radius. Then we recalculate
the recoil-structure correction with the obtained value of the proton magnetic
radius and take into account the radiative and binding contributions
to the proton-size term. At last we again find the proton-size
correction and extract the magnetic radius.

At first iteration we have calculated the recoil-structure
correction (integrals VO, VV, $\kappa^2$, No.1, and No.2
of Ref. \cite{bodwin1988}) with the new proton charge radius
$\la r^2\ra^{1/2}_{\rm E}=0.8750(68)$ fm \cite{mohr2004}
and with the same value for the magnetic radius of the proton.
The total recoil correction turns out to be $5.84$ ppm.
Accordingly, the proton-size correction
$\delta^{{\rm ps}}$ is given by
\be \label{ps}
\delta^{{\rm ps}}=\Delta E_{{\rm exp}}/E_{{\rm F}}-1
-\delta^{{\rm Dirac}}-\delta^{{\rm QED}}-\delta^{{\rm recoil}}
-\delta^{{\rm pol}}-\delta^{\mu{\rm vp}}
-\delta^{{\rm hvp}}-\delta^{{\rm weak}}\,.
\ee
This equation yields the  value
$\delta^{{\rm ps}}=-39.98(61)\times 10^{-6}$. The next section is devoted
to the deduction of the magnetic radius of the proton from
the proton-size correction and to the next iteraction.

\section{Magnetic radius of the proton}

The  proton-size correction has been first evaluated
in the non-relativistic limit by Zemach \cite{zemach1956}:
\be \label{ps_0}
\delta^{{\rm Zemach}}=-2\alpha Z\,\frac{m_{\rm e}m_{\rm p}}
{m_{\rm e}+m_{\rm p}}\, R_{\rm p}\,,
\ee
where
\be
R_{\rm p}=\int {\rm d}^3r\,{\rm d}^3r'\;\rho_{\rm E}({\bf r})
\rho_{\rm M}({\bf r}\pr)|{\bf r}-{\bf r}\pr|=
-\frac{4}{\pi}\int_0^\infty\frac{{\rm d}Q}{Q^2}\Bigl[\frac{\mu_{\rm N}}{\mu_{\rm p}}
\,G_{{\rm E}}(Q^2)\, G_{{\rm M}}(Q^2)-1\Bigr]
\ee
defines the Zemach radius of the proton.
Here $\rho_{{\rm E}}({\bf r})$ and $\rho_{{\rm M}}({\bf r})$ denote the
nuclear charge and magnetization distribution, respectively, where both densities
are normalized to unity.
The quantities $G_{{\rm E/M}}$ represent the electric and magnetic
form factors, respectively. Since we deal here only with the static
limit $(Q^0=0)$, we define them to be dependent only on the spatial
momentum transfer (squared), $Q^2>0$.
The charge and magnetic mean-square radii are defined by the formula
\be
\la r^2\ra_{\rm E/M}=\int {\rm d}^3r\, r^2\rho_{\rm E/M}({\bf r})\,.
\ee
In a first step we can approximate the  proton-size correction by the
Zemach formula. Then one can easily find the Zemach radius
$R_{\rm p}=1.058$ fm. The usual experimental fit for the proton
form factors is the dipole parametrization
\be
G_{\rm D}(Q^2)=\frac{1}{[1+Q^2R_{\rm D}^2]^2}\,,
\ee
which comes from the exponential model of the charge/magnetization distribution.
But recently, it was obtained in Jefferson Laboratory (JLab)
\cite{jones2000,gayou2001,gayou2002,guichon2003}, that
for $Q\ge 1$ GeV the behavior of the electric form factor differs from
the dipole parametrization. However, the Zemach correction is not sensitive
to the form factors behavior for $Q>0.8$ GeV. As it was shown in Ref.
\cite{brodsky2004}, the contribution to the Zemach correction
from the region $Q>0.8$ GeV is the same for different experimental models
of the proton electric and magnetic form factors. Therefore,
in what follows, we use the
dipole parametrizations for the form factors. The parameters $R_{\rm D}$,
 one for the
electric and another for the magnetic form factor, are directly connected with the
corresponding values of the root-mean-square radii. For the charge radius of the
proton we take the value $\la r^2\ra^{1/2}_{\rm E}=0.8750(68)$ fm \cite{mohr2004}
obtained from the latest comparison of the theoretical and experimental values of the
Lamb shift in hydrogen. Fixing the charge radius, we fit the magnetic radius such as
to reproduce the Zemach radius. As the result we find a preliminary magnetic
radius of the proton: $\la r^2\ra^{1/2}_{\rm M}=0.800$ fm. In order to estimate
the error associated with the model dependence we consider also the model
\be
G_{\rm JLab}(Q^2)=\left(1-0.13\frac{Q^2}{{\rm GeV}^2}\right)G_{\rm D}(Q^2)\,,
\ee
which is known as JLab model \cite{gayou2001}. The error appeared is on the level
of about $0.75$\%.

In a second step we account for corrections to the Zemach formula
\be
\delta^{\rm ps}=\delta^{\rm Zemach}+\delta^{\rm radiative}
+\delta^{\rm relativistic}\,,
\ee
where $\delta^{\rm radiative}$ is the radiative structure-dependent
correction obtained in \cite{karshenboim1997} and $\delta^{\rm relativistic}$
is the binding correction derived in \cite{volotka2003}. The radiative
correction has been derived assuming
the exponential model with the same parameter
$R_{\rm D}$ for both charge and magnetization
distributions, i.e.
\be
\delta^{\rm radiative}=-\delta^{\rm Zemach}
\frac{\alpha}{3\pi}\Bigl[
4\log{(m_{\rm e}R_{\rm D})}+\frac{4111}{420}\Bigr]\,.
\ee
The accuracy of this approximation is sufficient for our purpose.
Calculating $\delta^{\rm radiative}$ for different $R_{\rm D}$,
we obtain  $\delta^{\rm radiative} = 0.0153(2)\times \delta^{\rm Zemach}$.
The binding correction has been  expressed in terms of electric and
magnetic moments of the proton:
\be
\delta^{\rm relativistic}&=&
\delta^{\rm Zemach}(\alpha Z)^2
\Bigl[\frac74-\gamma-\log{(2\alpha Z)}\Bigr]
-2(\alpha Z)^3m_{\rm e}\la r\ra_{\rm E}\Biggl(
\frac{\la r\log{(m_{\rm e}r)}\ra_{\rm E}}{\la r\ra_{\rm E}}-
\frac{839}{750}\Biggr)\nonumber \\
&&-\frac{(\alpha Z)^3m_{\rm e}R_0}{5}\Biggl(
\frac{3\la r^4\ra_{\rm M}}{2R_0^4}
-\frac{19\la r^6\ra_{\rm M}}{42R_0^6}
+\frac{19\la r^8\ra_{\rm M}}{360R_0^8}
-\frac{2}{825}\frac{\la r^{10}\ra_{\rm M}}{R_0^{10}}\Biggr)\nonumber \\
&&-(\alpha Z)^3m_{\rm e}R_0\Biggl(\frac{\la r^2\ra_{\rm M}}{R_0^2}
-\frac{1}{10}\frac{\la r^4\ra_{\rm M}}{R_0^4}\Biggr)
\Biggl(\log{(m_{\rm e}R_0)}+\frac{1}{30}\Biggr)\,,
\ee
where $\gamma$ is Euler's constant and
$\la r^n\ra_{\rm E/M}=\int {\rm d}^3r\ r^n\rho_{\rm E/M}(r)$.
In part, this equation has been derived for the homogeneously charged sphere
model for the proton charge distribution
(with $R_0=\sqrt{5/3}\la r^2\ra_{\rm E}^{1/2}$).
Nevertheless, the error induced by using this formula in comparison with
other models for the charge distribution does not exceed $5$\%.
Employing the exponential model
for the charge and magnetization distributions with
electric and magnetic radii, $\la r^2\ra_{\rm E}^{1/2}=0.875$ fm
and $\la r^2\ra_{\rm M}^{1/2}=0.800$ fm, respectively,
we obtain $\delta^{\rm relativistic}=
0.0002\times\delta^{\rm Zemach}+1.4\times 10^{-8}$.
Thus the  proton-size correction takes the form
\be
\delta^{\rm ps}=1.0154(2)\times\delta^{\rm Zemach}+1.4\times 10^{-8}\,.
\ee
In addition, we need to correct the dominant term of the recoil contribution
with the magnetic radius $\la r^2\ra_{\rm M}^{1/2}=0.800$ fm.
As a result, the recoil correction turns out to be $5.94(6)$ ppm.

Deducing again the  proton-size correction by means of  equation (\ref{ps}),
we find the Zemach radius of the proton:
\begin{eqnarray}
R_{\rm p}=1.045(16) {\rm fm}\,.
\end{eqnarray}
This value differs from the one
obtained in \cite{dupays2003}, $R_{\rm p}=1.037(16)$ fm,
mainly due to the recalculated recoil corrections with
the new more precise charge and magnetic moment distributions.

\section{Discussion}

The value for the Zemach radius obtained above
enables us to determine an improved magnetic radius of the
proton:
\begin{eqnarray}
\la r^2\ra_{\rm M}^{1/2}=0.778(29) {\rm fm}\,.
\end{eqnarray}
The corresponding uncertainty is mainly due to errors associated with
the polarizability effect as well as the uncertainty of
the charge radius of the proton. In Table I we present the
final values
for the contributions to the hyperfine splitting in hydrogen.
The value for $\delta^{{\rm ps}}$ has been obtained
by means of the experimental energy splitting, according to equation (\ref{ps}).

Another way to determine the proton magnetic radius is based on
experimental data from  elastic electron-proton scattering.
Accordingly, Friar and Sick have recently determined the Zemach
radius of the proton to be $R_{\rm p}=1.086(12)$ fm \cite{friar2004} and
the proton charge radius $\la r^2\ra_{\rm E}^{1/2}=0.895(18)$ fm \cite{sick2003}.
Based on these values for $R_{\rm p}$ and $\la r^2\ra_{\rm E}^{1/2}$ and
employing an exponential parametrization for both electric and magnetic
form factors, we find the value of the proton magnetic radius
$\la r^2\ra_{\rm M}^{1/2}=0.824(27)$ fm. This value is in a good
agreement with the recent experimental value of Sick - $0.855(35)$ fm
presented in \cite{hammer2004}, and also the result of Hammer and
Mei\ss ner $0.857$ fm \cite{hammer2004} is not far away.

Recently Brodsky {\it et al.} \cite{brodsky2004} proposed another
method to extract the Zemach radius. They considered a rescaled
difference between the hyperfine splittings in hydrogen and muonium
\be\label{rescaled}
\frac{\Delta E^{{\rm p}}_{{\rm exp}}/E^{{\rm p}}_{{\rm F}}}
{\Delta E^\mu_{{\rm exp}}/E^\mu_{{\rm F}}}
=1+\delta^{{\rm hfs}}\,,
\ee
where p and $\mu$ indicate quantities which refer to the proton and muon,
respectively, $\Delta E^\mu_{{\rm exp}}=4\ 463.302\ 765(53)$ MHz
\cite{liu1999}, and $E^\mu_{{\rm F}}$ is the Fermi energy for muonium.
Employing recent values of the fundamental constants \cite{mohr2004}
they have obtained $\delta^{{\rm hfs}}=145.51$ ppm \cite{brodsky2004}.
The ground state hyperfine splitting in muonium can be written as
\be\label{muonium}
\Delta E^\mu_{{\rm theor}}=E^\mu_{{\rm F}}
(1+\delta^{{\rm Dirac}}+\delta^{{\rm QED}}+\delta_\mu^{{\rm recoil}}
  +\delta_\mu^{{\rm hvp}}+\delta_\mu^{{\rm weak}})\,.
\ee
The corrections $\delta^{{\rm Dirac}}$ and $\delta^{{\rm QED}}$ are
the same as in the case of hydrogen, $\delta_\mu^{{\rm hvp}}$
and $\delta_\mu^{{\rm weak}}$ are the hadronic vacuum-polarization
and the weak interaction contributions, respectively, and 
$\delta_\mu^{{\rm recoil}}$ is the recoil term, which consists of
relativistic and radiative parts.
From this formula together with Eqs. (\ref{hydrogen}) and (\ref{rescaled})
one can immediately derive the proton structure correction
\be\label{structure}
\delta_{{\rm p}}^{{\rm structure}}=
\delta^{{\rm hfs}}+\delta_\mu^{{\rm recoil}}
+\delta_\mu^{{\rm hvp}}+\delta_\mu^{{\rm weak}}
+\delta^{{\rm hfs}}(\delta^{{\rm Dirac}}+\delta^{{\rm QED}}
+\delta_\mu^{{\rm recoil}}+\delta_\mu^{{\rm hvp}}
+\delta_\mu^{{\rm weak}})\,.
\ee
If, following to Ref. \cite{brodsky2004},
we take into account only the relativistic part of the recoil
correction \cite{bodwin1988,karshenboim1993,kinoshita1998,melnikov2001,hill2001}
and neglect the contributions
$\delta_\mu^{{\rm hvp}}$ and $\delta_\mu^{{\rm weak}}$,
we obtain $\delta_{{\rm p}}^{{\rm structure}}=-31.8$ ppm.
This yields the Zemach radius 
 $R_{\rm p}=1.019(16)$ fm  \cite{brodsky2004}
that differs
significantly from our result, $R_{\rm p}=1.045(16)$ fm.
We have found, however, that the difference disappears if one includes
the omitted terms. This is mainly due to the radiative-recoil
correction evaluated in \cite{caswell1978,terray1982,
kinoshita1998,karshenboim1993,melnikov2001,hill2001,eides2004}.
With this term included, the total recoil correction is determined as
$\delta_\mu^{{\rm recoil}}=-178.33$ ppm.
The hadronic vacuum-polarization contribution obtained in
\cite{faustov1999,martynenko2004} is
$\delta_\mu^{{\rm hvp}}=0.05$ ppm, while the value of the
correction due to $Z^0$-boson exchange yields
$\delta_\mu^{{\rm weak}}=-0.01$ ppm \cite{beg1974,eides1996}.
Substituting these values into expression (\ref{structure}),
we find $\delta_{{\rm p}}^{{\rm structure}}=-32.64$ ppm.
Utilizing the values presented in Table I, we obtain for
the proton-size correction $\delta^{{\rm ps}}=-40.15$ ppm.
Then the Zemach radius can be easily determined with the result
$R_{\rm p}=1.047(16)$ fm, which is very close to the value
obtained in this work, $R_{\rm p}=1.045(16)$ fm.

As one can see, only a disagreement with the value for the Zemach radius and,
therefore, with this for the magnetic radius, as is obtained from the
electron-proton scattering experiments remains.
At present we have no explanation for this deviation.
One may hope, that a new determination of the proton charge radius via the
Lamb shift experiment with muonic hydrogen, which is now in progress
at PSI (Paul Scherrer Institute) \cite{pohl2001}, will elucidate
the situation.  From the theoretical point of view, an independent calculation
of the proton polarizability effect would be also desirable.

\section{Acknowledgements}

Valuable discussions with D. A. Glazov, S. G. Karshenboim,
A. P. Martynenko, and K. Pachucki are gratefully acknowledged.
This work was supported in parts
by the Russian Ministry of Education (grants no. A03-2.9-220, E02-31-49),
by RFBR (Grant No. 04-02-17574). G.P. and G.S. acknowledge
financial support by the BMBF, DAAD, DFG, and GSI.
The work of V.M.S. was  supported by the Alexander von Humboldt
Stiftung.

\newpage
\begin{table}
\caption{Numerical values for various corrections to the hyperfine splitting
in hydrogen together with the assigned errors.
The energies $\Delta E_{{\rm exp}}$ and
$E_{{\rm F}}$ are given in units of MHz.}
\begin{tabular}{llll}
                             &\hspace{1cm} Value   & \hspace{0.5cm} Error
                             & \hspace{1cm} Ref.          \\  \hline
$\Delta E_{{\rm exp}}$       &\ \ 1 420.405 751 767 & \ \ 0.000 000 001&\cite{hellwig1970}     \\
$E_{{\rm F}}$                &\ \ 1 418.840 08      & \ \ 0.000 02     &\cite{mohr2004}        \\
$\Delta E_{{\rm exp}}$/$E_{{\rm F}}$
                             &\ \ 1.001 103 49     & \ \ 0.000 000 01  &                       \\  \hline
$\delta^{{\rm Dirac}}$       &\ \ 0.000 079 88     &                   & \cite{breit1930}      \\
$\delta^{{\rm QED}}$         &\ \ 0.001 056 21     & \ \ 0.000 000 001 &
   \cite{pachucki1996,kinoshita1996,kinoshita1998,karshenboim2000,yerokhin2001,karshenboim2002}\\
$\delta^{{\rm ps}}$          &$-$ 0.000 040 11     & \ \ 0.000 000 61  &                       \\
$\delta^{{\rm recoil}}$      &\ \ 0.000 005 97     & \ \ 0.000 000 06  &
                                                   \cite{bodwin1988,karshenboim1997}, this work\\
$\delta^{{\rm pol}}$
                             &\ \ 0.000 001 4      & \ \ 0.000 000 6   & \cite{faustov2002}    \\
$\delta^{\mu{\rm vp}}$       &\ \ 0.000 000 07     & \ \ 0.000 000 02  & \cite{karshenboim1997}\\
$\delta^{{\rm hvp}}$         &\ \ 0.000 000 01     &              &
                                                              \cite{faustov1999,martynenko2004}\\
$\delta^{{\rm weak}}$        &\ \ 0.000 000 06     &              &
                                                                       \cite{beg1974,eides1996}\\ \hline
\end{tabular}
\end{table}


\begin{thebibliography}{45}
\expandafter\ifx\csname natexlab\endcsname\relax\def\natexlab#1{#1}\fi
\expandafter\ifx\csname bibnamefont\endcsname\relax
  \def\bibnamefont#1{#1}\fi
\expandafter\ifx\csname bibfnamefont\endcsname\relax
  \def\bibfnamefont#1{#1}\fi
\expandafter\ifx\csname citenamefont\endcsname\relax
  \def\citenamefont#1{#1}\fi
\expandafter\ifx\csname url\endcsname\relax
  \def\url#1{\texttt{#1}}\fi
\expandafter\ifx\csname urlprefix\endcsname\relax\def\urlprefix{URL }\fi
\providecommand{\bibinfo}[2]{#2}
\providecommand{\eprint}[2][]{\url{#2}}

\bibitem[{\citenamefont{Udem et~al.}(1997)\citenamefont{Udem, Huber, Gross,
  Reichert, Prevedelli, Weitz, and H{\"a}nsch}}]{udem1997}
\bibinfo{author}{\bibfnamefont{T.}~\bibnamefont{Udem}},
  \bibinfo{author}{\bibfnamefont{A.}~\bibnamefont{Huber}},
  \bibinfo{author}{\bibfnamefont{B.}~\bibnamefont{Gross}},
  \bibinfo{author}{\bibfnamefont{J.}~\bibnamefont{Reichert}},
  \bibinfo{author}{\bibfnamefont{M.}~\bibnamefont{Prevedelli}},
  \bibinfo{author}{\bibfnamefont{M.}~\bibnamefont{Weitz}}, \bibnamefont{and}
  \bibinfo{author}{\bibfnamefont{T.~W.} \bibnamefont{H{\"a}nsch}},
  \bibinfo{journal}{Phys. Rev. Lett.} \textbf{\bibinfo{volume}{79}},
  \bibinfo{pages}{2646} (\bibinfo{year}{1997}).

\bibitem[{\citenamefont{Schwob et~al.}(1999)\citenamefont{Schwob, Jozefowski,
  de~Beauvoir, Hilico, Nez, Julien, and Biraben}}]{schwob1999}
\bibinfo{author}{\bibfnamefont{C.}~\bibnamefont{Schwob}},
  \bibinfo{author}{\bibfnamefont{L.}~\bibnamefont{Jozefowski}},
  \bibinfo{author}{\bibfnamefont{B.}~\bibnamefont{de~Beauvoir}},
  \bibinfo{author}{\bibfnamefont{L.}~\bibnamefont{Hilico}},
  \bibinfo{author}{\bibfnamefont{F.}~\bibnamefont{Nez}},
  \bibinfo{author}{\bibfnamefont{L.}~\bibnamefont{Julien}}, \bibnamefont{and}
  \bibinfo{author}{\bibfnamefont{F.}~\bibnamefont{Biraben}},
  \bibinfo{journal}{Phys. Rev. Lett.} \textbf{\bibinfo{volume}{82}},
  \bibinfo{pages}{4960} (\bibinfo{year}{1999}).

\bibitem[{\citenamefont{de~Beauvoir et~al.}(2000)\citenamefont{de~Beauvoir,
  Schwob, Acef, Jozefowski, Hilico, Nez, Julien, Clairon, and
  Biraben}}]{beauvoir2000}
\bibinfo{author}{\bibfnamefont{B.}~\bibnamefont{de~Beauvoir}},
  \bibinfo{author}{\bibfnamefont{C.}~\bibnamefont{Schwob}},
  \bibinfo{author}{\bibfnamefont{O.}~\bibnamefont{Acef}},
  \bibinfo{author}{\bibfnamefont{L.}~\bibnamefont{Jozefowski}},
  \bibinfo{author}{\bibfnamefont{L.}~\bibnamefont{Hilico}},
  \bibinfo{author}{\bibfnamefont{F.}~\bibnamefont{Nez}},
  \bibinfo{author}{\bibfnamefont{L.}~\bibnamefont{Julien}},
  \bibinfo{author}{\bibfnamefont{A.}~\bibnamefont{Clairon}}, \bibnamefont{and}
  \bibinfo{author}{\bibfnamefont{F.}~\bibnamefont{Biraben}},
  \bibinfo{journal}{Eur. Phys. J. D} \textbf{\bibinfo{volume}{12}},
  \bibinfo{pages}{61} (\bibinfo{year}{2000}).

\bibitem[{\citenamefont{Pachucki and Jentschura}(2003)}]{pachucki2003}
\bibinfo{author}{\bibfnamefont{K.}~\bibnamefont{Pachucki}} \bibnamefont{and}
  \bibinfo{author}{\bibfnamefont{U.~D.} \bibnamefont{Jentschura}},
  \bibinfo{journal}{Phys. Rev. Lett.} \textbf{\bibinfo{volume}{91}},
  \bibinfo{pages}{113005} (\bibinfo{year}{2003}).

\bibitem[{\citenamefont{Verd{\'u} et~al.}(2004)\citenamefont{Verd{\'u},
  Djeki{\'c}, Stahl, Valenzuela, Vogel, Werth, Beier, Kluge, and
  Quint}}]{verdu2004}
\bibinfo{author}{\bibfnamefont{J.}~\bibnamefont{Verd{\'u}}},
  \bibinfo{author}{\bibfnamefont{S.}~\bibnamefont{Djeki{\'c}}},
  \bibinfo{author}{\bibfnamefont{S.}~\bibnamefont{Stahl}},
  \bibinfo{author}{\bibfnamefont{T.}~\bibnamefont{Valenzuela}},
  \bibinfo{author}{\bibfnamefont{M.}~\bibnamefont{Vogel}},
  \bibinfo{author}{\bibfnamefont{G.}~\bibnamefont{Werth}},
  \bibinfo{author}{\bibfnamefont{T.}~\bibnamefont{Beier}},
  \bibinfo{author}{\bibfnamefont{H.-J.} \bibnamefont{Kluge}}, \bibnamefont{and}
  \bibinfo{author}{\bibfnamefont{W.}~\bibnamefont{Quint}},
  \bibinfo{journal}{Phys. Rev. Lett.} \textbf{\bibinfo{volume}{92}},
  \bibinfo{pages}{093002} (\bibinfo{year}{2004}).

\bibitem[{\citenamefont{Mohr and Taylor}(2004)}]{mohr2004}
\bibinfo{author}{\bibfnamefont{P.~J.} \bibnamefont{Mohr}} \bibnamefont{and}
  \bibinfo{author}{\bibfnamefont{B.~N.} \bibnamefont{Taylor}},
  \bibinfo{journal}{Rev. Mod. Phys.} \textbf{\bibinfo{volume}{76}},
  \bibinfo{pages}{to be published} (\bibinfo{year}{2004}).

\bibitem[{\citenamefont{Melnikov and van Ritbergen}(2000)}]{melnikov2000}
\bibinfo{author}{\bibfnamefont{K.}~\bibnamefont{Melnikov}} \bibnamefont{and}
  \bibinfo{author}{\bibfnamefont{T.}~\bibnamefont{van Ritbergen}},
  \bibinfo{journal}{Phys. Rev. Lett.} \textbf{\bibinfo{volume}{84}},
  \bibinfo{pages}{1673} (\bibinfo{year}{2000}).

\bibitem[{\citenamefont{Pachucki}(2001)}]{pachucki2001}
\bibinfo{author}{\bibfnamefont{K.}~\bibnamefont{Pachucki}},
  \bibinfo{journal}{Phys. Rev. A} \textbf{\bibinfo{volume}{63}},
  \bibinfo{pages}{042503} (\bibinfo{year}{2001}).

\bibitem[{\citenamefont{Karshenboim}(Springer, Berlin, 2003)}]{karshenboim2003}
\bibinfo{author}{\bibfnamefont{S.~G.} \bibnamefont{Karshenboim}},
  \bibinfo{journal}{Precision physics of simple atomic systems, ed. by S. G.
  Karshenboim and V. B. Smirnov} p. \bibinfo{pages}{141}
  (\bibinfo{year}{Springer, Berlin, 2003}).

\bibitem[{\citenamefont{Hellwig et~al.}(1970)\citenamefont{Hellwig, Vessot,
  Levine, Zitzewitz, Allan, and Glaze}}]{hellwig1970}
\bibinfo{author}{\bibfnamefont{H.}~\bibnamefont{Hellwig}},
  \bibinfo{author}{\bibfnamefont{R.~F.~C.} \bibnamefont{Vessot}},
  \bibinfo{author}{\bibfnamefont{M.~W.} \bibnamefont{Levine}},
  \bibinfo{author}{\bibfnamefont{P.~W.} \bibnamefont{Zitzewitz}},
  \bibinfo{author}{\bibfnamefont{D.~W.} \bibnamefont{Allan}}, \bibnamefont{and}
  \bibinfo{author}{\bibfnamefont{D.~J.} \bibnamefont{Glaze}},
  \bibinfo{journal}{IEEE Trans. Instr. Meas. IM} \textbf{\bibinfo{volume}{19}},
  \bibinfo{pages}{200} (\bibinfo{year}{1970}).

\bibitem[{\citenamefont{Dupays et~al.}(2003)\citenamefont{Dupays, Beswick,
  Lepetit, Rizzo, and Bakalov}}]{dupays2003}
\bibinfo{author}{\bibfnamefont{A.}~\bibnamefont{Dupays}},
  \bibinfo{author}{\bibfnamefont{A.}~\bibnamefont{Beswick}},
  \bibinfo{author}{\bibfnamefont{B.}~\bibnamefont{Lepetit}},
  \bibinfo{author}{\bibfnamefont{C.}~\bibnamefont{Rizzo}}, \bibnamefont{and}
  \bibinfo{author}{\bibfnamefont{D.}~\bibnamefont{Bakalov}},
  \bibinfo{journal}{Phys. Rev. A} \textbf{\bibinfo{volume}{68}},
  \bibinfo{pages}{052503} (\bibinfo{year}{2003}).

\bibitem[{\citenamefont{Sick}(2003)}]{sick2003}
\bibinfo{author}{\bibfnamefont{I.}~\bibnamefont{Sick}}, \bibinfo{journal}{Phys.
  Lett. B} \textbf{\bibinfo{volume}{576}}, \bibinfo{pages}{62}
  (\bibinfo{year}{2003}).

\bibitem[{\citenamefont{Friar and Sick}(2004)}]{friar2004}
\bibinfo{author}{\bibfnamefont{J.~L.} \bibnamefont{Friar}} \bibnamefont{and}
  \bibinfo{author}{\bibfnamefont{I.}~\bibnamefont{Sick}},
  \bibinfo{journal}{Phys. Lett. B} \textbf{\bibinfo{volume}{579}},
  \bibinfo{pages}{285} (\bibinfo{year}{2004}).

\bibitem[{\citenamefont{Hammer and Mei{\ss}ner}(2004)}]{hammer2004}
\bibinfo{author}{\bibfnamefont{H.-W.} \bibnamefont{Hammer}} \bibnamefont{and}
  \bibinfo{author}{\bibfnamefont{U.-G.} \bibnamefont{Mei{\ss}ner}},
  \bibinfo{journal}{Eur. Phys. J. A} \textbf{\bibinfo{volume}{20}},
  \bibinfo{pages}{469} (\bibinfo{year}{2004}).

\bibitem[{\citenamefont{{S. J. Brodsky, C. E. Carlson, J. R. Hiller, and D. S.
  Hwang, e-print hep-ph/0408131 (2004).}}()}]{brodsky2004}
\bibinfo{author}{\bibnamefont{{S. J. Brodsky, C. E. Carlson, J. R. Hiller, and
  D. S. Hwang, e-print hep-ph/0408131 (2004).}}}

\bibitem[{\citenamefont{Fermi}(1930)}]{fermi1930}
\bibinfo{author}{\bibfnamefont{E.}~\bibnamefont{Fermi}}, \bibinfo{journal}{Z.
  Phys.} \textbf{\bibinfo{volume}{60}}, \bibinfo{pages}{320}
  (\bibinfo{year}{1930}).

\bibitem[{\citenamefont{Breit}(1930)}]{breit1930}
\bibinfo{author}{\bibfnamefont{G.}~\bibnamefont{Breit}},
  \bibinfo{journal}{Phys. Rev.} \textbf{\bibinfo{volume}{35}},
  \bibinfo{pages}{1447} (\bibinfo{year}{1930}).

\bibitem[{\citenamefont{Pachucki}(1996)}]{pachucki1996}
\bibinfo{author}{\bibfnamefont{K.}~\bibnamefont{Pachucki}},
  \bibinfo{journal}{Phys. Rev. A} \textbf{\bibinfo{volume}{54}},
  \bibinfo{pages}{1994} (\bibinfo{year}{1996}).

\bibitem[{\citenamefont{Kinoshita and Nio}(1996)}]{kinoshita1996}
\bibinfo{author}{\bibfnamefont{T.}~\bibnamefont{Kinoshita}} \bibnamefont{and}
  \bibinfo{author}{\bibfnamefont{M.}~\bibnamefont{Nio}},
  \bibinfo{journal}{Phys. Rev. D} \textbf{\bibinfo{volume}{53}},
  \bibinfo{pages}{4909} (\bibinfo{year}{1996}).

\bibitem[{\citenamefont{Kinoshita}(1998)}]{kinoshita1998}
\bibinfo{author}{\bibfnamefont{T.}~\bibnamefont{Kinoshita}},
  \bibinfo{journal}{e-print hep-ph/9808351}  (\bibinfo{year}{1998}).

\bibitem[{\citenamefont{{S. G. Karshenboim, V. G. Ivanov, and V. M. Shabaev,
  JETP {\bf 90}, 59 (2000); Can. J. Phys. {\bf 76}, 503
  (1998)}}()}]{karshenboim2000}
\bibinfo{author}{\bibnamefont{{S. G. Karshenboim, V. G. Ivanov, and V. M.
  Shabaev, JETP {\bf 90}, 59 (2000); Can. J. Phys. {\bf 76}, 503 (1998)}}}.

\bibitem[{\citenamefont{Yerokhin and Shabaev}(2001)}]{yerokhin2001}
\bibinfo{author}{\bibfnamefont{V.~A.} \bibnamefont{Yerokhin}} \bibnamefont{and}
  \bibinfo{author}{\bibfnamefont{V.~M.} \bibnamefont{Shabaev}},
  \bibinfo{journal}{Phys. Rev. A} \textbf{\bibinfo{volume}{64}},
  \bibinfo{pages}{012506} (\bibinfo{year}{2001}).

\bibitem[{\citenamefont{Karshenboim and Ivanov}(2002)}]{karshenboim2002}
\bibinfo{author}{\bibfnamefont{S.~G.} \bibnamefont{Karshenboim}}
  \bibnamefont{and} \bibinfo{author}{\bibfnamefont{V.~G.}
  \bibnamefont{Ivanov}}, \bibinfo{journal}{Eur. Phys. J. D}
  \textbf{\bibinfo{volume}{19}}, \bibinfo{pages}{13} (\bibinfo{year}{2002}).

\bibitem[{\citenamefont{{R. N. Faustov and A. P. Martynenko, Phys. At. Nucl.
  {\bf 65}, 265 (2002); Yad. Fiz. {\bf 65}, 291 (2002); e-print hep-ph/0007044
  (2000).}}()}]{faustov2002}
\bibinfo{author}{\bibnamefont{{R. N. Faustov and A. P. Martynenko, Phys. At.
  Nucl. {\bf 65}, 265 (2002); Yad. Fiz. {\bf 65}, 291 (2002); e-print
  hep-ph/0007044 (2000).}}}

\bibitem[{\citenamefont{Karshenboim}(1997)}]{karshenboim1997}
\bibinfo{author}{\bibfnamefont{S.~G.} \bibnamefont{Karshenboim}},
  \bibinfo{journal}{Phys. Lett. A} \textbf{\bibinfo{volume}{225}},
  \bibinfo{pages}{97} (\bibinfo{year}{1997}).

\bibitem[{\citenamefont{{R. N. Faustov, A. Karimkhodzhaev, and A. P.
  Martynenko, Phys. Rev. A {\bf 59}, 2498 (1999); Phys. At. Nucl. {\bf 62},
  2103 (1999); Yad. Fiz. {\bf 62}, 2284 (1999); e-print hep-ph/9808365
  (1998).}}()}]{faustov1999}
\bibinfo{author}{\bibnamefont{{R. N. Faustov, A. Karimkhodzhaev, and A. P.
  Martynenko, Phys. Rev. A {\bf 59}, 2498 (1999); Phys. At. Nucl. {\bf 62},
  2103 (1999); Yad. Fiz. {\bf 62}, 2284 (1999); e-print hep-ph/9808365
  (1998).}}}

\bibitem[{\citenamefont{{A. P. Martynenko and R. N. Faustov, JETP {\bf 98}, 39
  (2004); e-print hep-ph/0312116 (2004).}}()}]{martynenko2004}
\bibinfo{author}{\bibnamefont{{A. P. Martynenko and R. N. Faustov, JETP {\bf
  98}, 39 (2004); e-print hep-ph/0312116 (2004).}}}

\bibitem[{\citenamefont{{M. A. B. B{\'e}g and G. Feinberg, Phys. Rev. Lett.
  {\bf 33}, 606 (1974); {\bf 35}, 130 (1975).}}()}]{beg1974}
\bibinfo{author}{\bibnamefont{{M. A. B. B{\'e}g and G. Feinberg, Phys. Rev.
  Lett. {\bf 33}, 606 (1974); {\bf 35}, 130 (1975).}}}

\bibitem[{\citenamefont{Eides}(1996)}]{eides1996}
\bibinfo{author}{\bibfnamefont{M.~I.} \bibnamefont{Eides}},
  \bibinfo{journal}{Phys. Rev. A} \textbf{\bibinfo{volume}{53}},
  \bibinfo{pages}{2953} (\bibinfo{year}{1996}).

\bibitem[{\citenamefont{Pineda}(2003)}]{pineda2003}
\bibinfo{author}{\bibfnamefont{A.}~\bibnamefont{Pineda}},
  \bibinfo{journal}{Phys. Rev. C} \textbf{\bibinfo{volume}{67}},
  \bibinfo{pages}{025201} (\bibinfo{year}{2003}).

\bibitem[{\citenamefont{Bodwin and Yennie}(1988)}]{bodwin1988}
\bibinfo{author}{\bibfnamefont{G.~T.} \bibnamefont{Bodwin}} \bibnamefont{and}
  \bibinfo{author}{\bibfnamefont{D.~R.} \bibnamefont{Yennie}},
  \bibinfo{journal}{Phys. Rev. D} \textbf{\bibinfo{volume}{37}},
  \bibinfo{pages}{498} (\bibinfo{year}{1988}).

\bibitem[{\citenamefont{Zemach}(1956)}]{zemach1956}
\bibinfo{author}{\bibfnamefont{A.~C.} \bibnamefont{Zemach}},
  \bibinfo{journal}{Phys. Rev.} \textbf{\bibinfo{volume}{104}},
  \bibinfo{pages}{1771} (\bibinfo{year}{1956}).

\bibitem[{\citenamefont{{M. K. Jones, K. A. Aniol, F. T. Baker, J. Berthot, P.
  Y. Bertin, W. Bertozzi, A. Besson, et al.}}(2000)}]{jones2000}
\bibinfo{author}{\bibnamefont{{M. K. Jones, K. A. Aniol, F. T. Baker, J.
  Berthot, P. Y. Bertin, W. Bertozzi, A. Besson, et al.}}},
  \bibinfo{journal}{Phys. Rev. Lett.} \textbf{\bibinfo{volume}{84}},
  \bibinfo{pages}{1398} (\bibinfo{year}{2000}).

\bibitem[{\citenamefont{{O. Gayou, K. Wijesooriya, A. Afanasev, M. Amarian, K.
  Aniol, S. Becher, K. Benslama, et al.}}(2001)}]{gayou2001}
\bibinfo{author}{\bibnamefont{{O. Gayou, K. Wijesooriya, A. Afanasev, M.
  Amarian, K. Aniol, S. Becher, K. Benslama, et al.}}}, \bibinfo{journal}{Phys.
  Rev. C} \textbf{\bibinfo{volume}{64}}, \bibinfo{pages}{038202}
  (\bibinfo{year}{2001}).

\bibitem[{\citenamefont{{O. Gayou, K. A. Aniol, T. Averett, F. Benmokhtar, W.
  Bertozzi, L. Bimbot, E. J. Brash, et al.}}(2002)}]{gayou2002}
\bibinfo{author}{\bibnamefont{{O. Gayou, K. A. Aniol, T. Averett, F.
  Benmokhtar, W. Bertozzi, L. Bimbot, E. J. Brash, et al.}}},
  \bibinfo{journal}{Phys. Rev. Lett.} \textbf{\bibinfo{volume}{88}},
  \bibinfo{pages}{092301} (\bibinfo{year}{2002}).

\bibitem[{\citenamefont{Guichon and Vanderhaeghen}(2003)}]{guichon2003}
\bibinfo{author}{\bibfnamefont{P.~A.~M.} \bibnamefont{Guichon}}
  \bibnamefont{and}
  \bibinfo{author}{\bibfnamefont{M.}~\bibnamefont{Vanderhaeghen}},
  \bibinfo{journal}{Phys. Rev. Lett.} \textbf{\bibinfo{volume}{91}},
  \bibinfo{pages}{142303} (\bibinfo{year}{2003}).

\bibitem[{\citenamefont{Volotka et~al.}(2003)\citenamefont{Volotka, Shabaev,
  Plunien, and Soff}}]{volotka2003}
\bibinfo{author}{\bibfnamefont{A.~V.} \bibnamefont{Volotka}},
  \bibinfo{author}{\bibfnamefont{V.~M.} \bibnamefont{Shabaev}},
  \bibinfo{author}{\bibfnamefont{G.}~\bibnamefont{Plunien}}, \bibnamefont{and}
  \bibinfo{author}{\bibfnamefont{G.}~\bibnamefont{Soff}},
  \bibinfo{journal}{Eur. Phys. J. D} \textbf{\bibinfo{volume}{23}},
  \bibinfo{pages}{51} (\bibinfo{year}{2003}).

\bibitem[{\citenamefont{{W. Liu, M. G. Boshier, S. Dhawan, O. van Dyck, P.
  Egan, X. Fei, M. Grosse Perdekamp, et al.}}(1999)}]{liu1999}
\bibinfo{author}{\bibnamefont{{W. Liu, M. G. Boshier, S. Dhawan, O. van Dyck,
  P. Egan, X. Fei, M. Grosse Perdekamp, et al.}}}, \bibinfo{journal}{Phys. Rev.
  Lett.} \textbf{\bibinfo{volume}{82}}, \bibinfo{pages}{711}
  (\bibinfo{year}{1999}).

\bibitem[{\citenamefont{Karshenboim}(1993)}]{karshenboim1993}
\bibinfo{author}{\bibfnamefont{S.~G.} \bibnamefont{Karshenboim}},
  \bibinfo{journal}{JETP} \textbf{\bibinfo{volume}{76}}, \bibinfo{pages}{541}
  (\bibinfo{year}{1993}).

\bibitem[{\citenamefont{Melnikov and Yelkhovsky}(2001)}]{melnikov2001}
\bibinfo{author}{\bibfnamefont{K.}~\bibnamefont{Melnikov}} \bibnamefont{and}
  \bibinfo{author}{\bibfnamefont{A.}~\bibnamefont{Yelkhovsky}},
  \bibinfo{journal}{Phys. Rev. Lett.} \textbf{\bibinfo{volume}{86}},
  \bibinfo{pages}{1498} (\bibinfo{year}{2001}).

\bibitem[{\citenamefont{Hill}(2001)}]{hill2001}
\bibinfo{author}{\bibfnamefont{R.~J.} \bibnamefont{Hill}},
  \bibinfo{journal}{Phys. Rev. Lett.} \textbf{\bibinfo{volume}{86}},
  \bibinfo{pages}{3280} (\bibinfo{year}{2001}).

\bibitem[{\citenamefont{Caswell and Lepage}(1978)}]{caswell1978}
\bibinfo{author}{\bibfnamefont{W.~E.} \bibnamefont{Caswell}} \bibnamefont{and}
  \bibinfo{author}{\bibfnamefont{G.~P.} \bibnamefont{Lepage}},
  \bibinfo{journal}{Phys. Rev. Lett.} \textbf{\bibinfo{volume}{41}},
  \bibinfo{pages}{1092} (\bibinfo{year}{1978}).

\bibitem[{\citenamefont{Terray and Yennie}(1982)}]{terray1982}
\bibinfo{author}{\bibfnamefont{E.~A.} \bibnamefont{Terray}} \bibnamefont{and}
  \bibinfo{author}{\bibfnamefont{D.~R.} \bibnamefont{Yennie}},
  \bibinfo{journal}{Phys. Rev. Lett.} \textbf{\bibinfo{volume}{48}},
  \bibinfo{pages}{1803} (\bibinfo{year}{1982}).

\bibitem[{\citenamefont{{M. I. Eides, H. Grotch, and V. A. Shelyuto, e-print
  hep-ph/0412372 (2004).}}()}]{eides2004}
\bibinfo{author}{\bibnamefont{{M. I. Eides, H. Grotch, and V. A. Shelyuto,
  e-print hep-ph/0412372 (2004).}}}

\bibitem[{\citenamefont{Pohl et~al.}(Springer, Berlin, Heidelberg,
  2001)\citenamefont{Pohl, Biraben, Conde, Donche-Gay, H{\"a}nsch, Hartmann,
  Hauser, Hughes, Huot, Indelicato et~al.}}]{pohl2001}
\bibinfo{author}{\bibfnamefont{R.}~\bibnamefont{Pohl}},
  \bibinfo{author}{\bibfnamefont{F.}~\bibnamefont{Biraben}},
  \bibinfo{author}{\bibfnamefont{C.~A.~N.} \bibnamefont{Conde}},
  \bibinfo{author}{\bibfnamefont{C.}~\bibnamefont{Donche-Gay}},
  \bibinfo{author}{\bibfnamefont{T.~W.} \bibnamefont{H{\"a}nsch}},
  \bibinfo{author}{\bibfnamefont{F.~J.} \bibnamefont{Hartmann}},
  \bibinfo{author}{\bibfnamefont{P.}~\bibnamefont{Hauser}},
  \bibinfo{author}{\bibfnamefont{V.~W.} \bibnamefont{Hughes}},
  \bibinfo{author}{\bibfnamefont{O.}~\bibnamefont{Huot}},
  \bibinfo{author}{\bibfnamefont{P.}~\bibnamefont{Indelicato}},
  \bibnamefont{et~al.}, \bibinfo{journal}{Hydrogen atom: Precision physics of
  simple atomic systems, ed. by S. G. Karshenboim, et al.} p.
  \bibinfo{pages}{454} (\bibinfo{year}{Springer, Berlin, Heidelberg, 2001}).

\end{thebibliography}
\end{document}